\documentclass[preprints,article,accept,moreauthors,pdftex]{mdpi} 

\usepackage{amsfonts}

\renewcommand{\BibitemShut}[1]{}

\newcommand\cL{{\mathcal L}}
\newcommand\cM{{\mathcal M}}

\newcommand\sdot{\stackrel{.}{\sigma}}

\usepackage{verbatim}
\newcommand{\HCd}{\mathcal{H}}

\def\HCdt0{\tilde{\HCd}_{0}}

\newcommand\rf[1]{(\ref{eq:#1})}
\newcommand\lab[1]{\label{eq:#1}}
\newcommand\nonu{\nonumber}
\newcommand\br{\begin{eqnarray}}
\newcommand\er{\end{eqnarray}}
\newcommand\be{\begin{equation}}
\newcommand\ee{\end{equation}}

\newcommand\lb{\lbrack}
\newcommand\rb{\rbrack}

\renewcommand\({\left(}
\renewcommand\){\right)}

\newcommand\bc{\begin{center}}
\newcommand\ec{\end{center}}

\newcommand\partder[2]{\frac{{\partial {#1}}}{{\partial {#2}}}}

\renewcommand\a{\alpha}
\renewcommand\b{\beta}

\renewcommand\d{\delta}

\newcommand\vareps{\varepsilon}

\renewcommand\G{\Gamma}

\newcommand\h{\frac{1}{2}}
\renewcommand\k{\kappa}
\renewcommand\l{\lambda}
\renewcommand\L{\Lambda}
\newcommand\m{\mu}
\newcommand\n{\nu}
\newcommand\om{\omega}

\renewcommand\P{\Phi}
\newcommand\pa{\partial}

\newcommand\s{\sigma}
\renewcommand\S{\Sigma}

\newcommand\wti{\widetilde}
\newcommand\twomat[4]{\left(\begD{array}{cc}  
{#1} & {#2} \\ {#3} & {#4} \end{array} \right)}

\newcommand{\ct}[1]{\cite{#1}}

\newcommand\phidot{\stackrel{.}{\phi}}

\newcommand\adot{\stackrel{.}{a}}

\firstpage{1} 
\pubvolume{xx}
\issuenum{1}
\articlenumber{5}
\pubyear{2019}
\copyrightyear{2019}
\history{Received: date; Accepted: date; Published: date}

\Title{Quintessential Inflation with Dynamical Higgs Generation as an Affine Gravity}

\Author{D. Benisty$^{1,2}$, E. I. Guendelman$^{1,2,3}$, E. Nissimov*$^{4}$ and S. Pacheva$^{4}$}

\address{%
$^{1}$ \quad Physics Department, Ben-Gurion University of the Negev, Beer-Sheva 84105, Israel\\
$^{2}$ \quad Frankfurt Institute for Advanced Studies (FIAS), Ruth-Moufang-Strasse~1, 60438 Frankfurt am Main, Germany\\ 
$^{3}$ \quad Bahamas Advanced Study Institute and Conferences, 4A Ocean Heights, Hill View Circle, Stella Maris, Long Island, The Bahamas\\
$^{4}$ Institute for Nuclear Research and Nuclear Energy, Bulgarian Academy of Sciences, Sofia, Bulgaria}

\corres{Correspondence: nissimov@inrne.bas.bg}

\abstract{First, we propose a scale-invariant modified gravity 
interacting with a neutral scalar inflaton and a Higgs-like $SU(2)\times U(1)$ 
iso-doublet scalar field based on the formalism of non-Riemannian (metric-independent) 
spacetime volume-elements. 
This model describes in the physical Einstein frame a
quintessential inflationary scenario driven by the ``inflaton'' together with
gravity-inflaton assisted dynamical spontaneous $SU(2)\times U(1)$ symmetry
breaking in the post-inflationary universe, 
whereas $SU(2)\times U(1)$ symmetry remains intact in the inflationary epoch.
Next, we find the explicit representation of the latter quintessential 
inflationary model with a dynamical Higgs effect as an Eddington-type purely affine 
gravity.}


\begin{document}
\section{Introduction}

Studies in cosmology are dominated by the fundamental concept of
``inflation'' -- a period of exponential expansion, which provides a plausible solution
for the ``puzzles'' of
the Big-Bang cosmology (the horizon problem, the flatness problem, the
magnetic monopole problem, etc.) 
\ct{Guth:1980zm,Starobinsky:1979ty,Kazanas:1980tx,Starobinsky:1980te,Linde:1981mu,Albrecht:1982wi,Barrow:1983rx,Blau:1986cw}.
For more extensive accounts, see the books
\ct{early-univ-1,early-univ-2,early-univ-3,early-univ-4,early-univ-5,early-univ-6,primordial-1,primordial-2,lyth-2017,calcagni-2017,rubakov-2018,piatella-2018,Benisty:2018gzx}.
The most widely discussed  mechanism for generating a period of accelerated 
expansion is through the presence of some vacuum energy. In the context of models
with scalar field(s)-driven inflation, vacuum energy density appears naturally 
when the scalar field(s) acquire an effective potential $U_{\rm eff}$ which has 
flat regions so that the scalar field(s) can ``slowly roll''
\ct{slow-roll-1,slow-roll-2,slow-roll-param-1,slow-roll-param-2,Benisty:2019vej}
and its/their kinetic energy can be neglected resulting in an
energy-momentum tensor of the form $T_{\m\n} \simeq - g_{\m\n} U_{\rm eff}$.

With the discovery of the accelerating expansion of the present 
universe \ct{accel-exp-1,accel-exp-2,accel-exp-3,Saridakis:2018unr,Vasak:2019nmy,DiValentino:2019jae,Perez:2020cwa,Anagnostopoulos:2017iao,Struckmeier:2017vkf} it appears plausible that a small vacuum energy density, usually referred in this case as ``dark energy'', is also present even today. The two vacuum energy densities -- the one of inflation and the other of the dark energy dominated universe nowadays, have however a totally different scale which demans a plausible explanation of how cosmological evolution may naturally interpolate between such two apparently quite distinctive physical situations.

The possibility of continuously connecting an inflationary phase of the
``early'' universe to a slowly accelerating universe of nowadays through the 
evolution of a single scalar field -- the
{\em quintessential inflation scenario} -- has been first studied in 
 \ct{peebles-vilenkin}. Subsequently, a multitude of different
quintessential inflationary models have been proposed: (a) based on modified
$f(R)$ gravity \ct{odintsov-2003,odintsov-2008,DeLaurentis:2015fea}; (b) based on the 
k-essence concept \ct{k-essence-1,k-essence-2,k-essence-3,k-essence-4,saitou-nojiri}; 
based on the ``variable gravity'' model \ct{wetterich}.
For extensive list of references to earlier work on the topic of quintessential 
inflation, see
Refs.\cite{BuenoSanchez:2006epu,BuenoSanchez:2006fhh,Dimopoulos:2001ix,Bezrukov:2007ep,Choudhury:2013zna,Dimopoulos:2017tud,Dimopoulos:2017zvq,Choudhury:2017cos,Gundhi:2018wyz,Ema:2019fdd}
(some of them focusing on Higgs inflation) and
Refs.\ct{murzakulov-etal,Hossain:2014xha,Hossain:2014coa,Hossain:2014ova,Hossain:2014zma,Geng:2015fla,Geng:2017mic},
In particular, see the recent Ref.\ct{Kleidis:2019ywv} for quintessential inflation 
in the context of Einstein-Gauss-Bonnet gravity and Ref.\ct{Dimopoulos:2019gpz}
about warm quintessential inflation. 

Another parallel groundbreaking development alongside the quintessential
inflationary cosmology is the advent of extended modified gravitational theories.
The main motivation aims to overcome the limitations of the canonical Einstein's 
general relativity manifesting themselves in: (i) Cosmology -- 
for solving the problems of dark energy and dark matter and explaining the 
large scale structure of the Universe  
\ct{Perlmutter:1998np,Copeland:2006wr,Novikov:2016fzd};
(ii) Quantum field theory in curved spacetime -- because of the non-renormalizabilty 
of ultraviolet divergences in higher loops 
\ct{Benitez:2020szx,Budge:2020oyl,Bell:2020qus,Frohlich:2020igy,DAmbrosio:2020yfa,Novikov:2016hrc}; 
(iii) Modern string theory -- because of the natural appearance of higher-order 
curvature invariants and scalar-tensor couplings in low-energy effective field 
theories \ct{Dekens:2019ept,Ma:2019wbc,Jenkins:2017jig,Brandyshev:2017ywi,Gomez:2020xdb}.

Various classes of modified gravity theories have been employed to construct 
plausible inflationary models: $f(R)$-gravity, scalar-tensor gravity, 
Gauss-Bonnet gravity (see   \ct{Capozziello:2010zz,Nojiri:2017ncd} for 
extensive review); also recent proposals based on non-local gravity 
(\ct{Dimitrijevic:2019pct} and references therein) or based on brane-world scenarios 
(\ct{Bilic:2018uqx} and references therein). Let us recall 
the first early successful cosmological model based on the extended 
$f(R)= R + R^2$-gravity producing the  classical Starobinsky inflationary scalar field 
potential \ct{Starobinsky:1979ty}.

For a recent detailed work on quintessential inflation based on $f(R)$-gravity, where
the role of dark matter is being played by axions, see Ref.\ct{Odintsov:2020nwm}. 

A broad class of actively developed modified/extended gravitational theories is 
based on employing (one or more) alternative non-Riemannian spacetime volume-forms, 
\textsl{i.e.}, metric-independent generally covariant volume-elements in the 
pertinent Lagrangian actions on spacetime manifolds with an ordinary Riemannian 
geometry, instead of (or alongside with) the canonical Riemannian volume-element 
$\sqrt{-g} \, d^4 x$, whose density is given by the square-root of the 
determinant of the Riemannian metric $\sqrt{-g} \equiv \sqrt{-\det\Vert g_{\m\n}\Vert}$. 

Originally the formalism employing non-Riemannian volume-elements in
generally-covariant Lagrangian actions as in Eq.\rf{NRVF-0} below  
was proposed in \ct{TMT-orig-0,Hehl,TMT-orig-1,TMT-orig-2,5thforce}.
The concise geometric formulation was presented in \ct{susyssb-1,grav-bags}. A brief
outline of the basics of the formalism of non-Riemannian volume-elements is
given in Section 2 below.

This formalism was used as a basis for constructing a series of modified gravity-matter
models describing unified dark energy and dark matter scenario \ct{dusty,dusty-2},
quintessential cosmological models with gravity-assisted and inflaton-assisted
dynamical suppression (in the ``early'' universe) or dynamical generation (in the
post-inflationary universe) of electroweak spontaneous symmetry
breaking and charge confinement \ct{grf-essay,varna-17,bpu-10}, as well as a novel 
mechanism for the supersymmetric Brout-Englert-Higgs effect (dynamical
spontaneous supersymmetry breaking) in supergravity \ct{susyssb-1}.

In the present paper our first principal goal is to analyze (Section 3 below) the close
interplay between cosmological dynamics and the pattens of (spontaneous)
symmetry breaking along the history of universe, which itself is one of the
most important paradigms at the interface of particle physics and cosmology.
We will extend our construction, started in   \ct{grf-essay}, of a modified 
gravity model coupled to (the Higgs part) of the standard electroweak matter content
(see, \textsl{e.g.} \ct{standard-particle-model-1,standard-particle-model-2} 
besides the scalar ``inflaton'' field. The main
aim here is to provide an explicit realization from first (Lagrangian
action) principles of the remarkable proposal of Bekenstein
\ct{bekenstein-86} about the so called gravity-assisted dynamical generation
of the Higgs effect -- dynamical symmetry breaking of the electroweak 
$SU(2)\times U(1)$ symmetry -- without introducing unnatural (according to
Bekenstein's opinion) ingredients like negative (``ghost''-like) mass
squared and quartic self-interaction for the Higgs field. Here we study the
interrelation between the presence or absence of dynamical spontaneous electroweak 
symmetry breaking and the different stages of universe's evolution driven by
the ``inflaton'' -- triggering inflation in the ``early'' universe as well
as representing quintessential variable dark-energy in the ``late'' universe.

It is shown that during inflation there is no spontaneous electroweak
symmetry breaking and the Higgs field resides in its ``wrong'' vacuum state 
(``wrong'' from the point of view of standard high-energy particle physics).
The non-trivial symmetry-breaking Higgs vacuum is dynamically generated in
the post-inflationary epoch.

Let us specifically stress that this mechanism is different from the  widely 
discussed scenario of Higgs inflation,  
where the Higgs field triggers the inflation in the ``early'' universe through a 
non-minimal coupling to gravity 
\cite{Rubio:2018ogq,Antoniadis:2020dfq,Ouseph:2020kcz,Ema:2020zvg,Torabian:2020bku,Adshead:2020ijf,Okada:2020cvq,Shaposhnikov:2020fdv,Tenkanen:2020cvw,Shaposhnikov:2020geh,Barrie:2020hiu,Sato:2020ghj,Passaglia:2019ueo,Mishra:2019ymr,Gialamas:2019nly,Benisty:2019pxb}.
In our scenario the impact of the Higgs field dynamics starts after end of inflation. 

Another ground-laying branch of gravitational theories is 
the purely affine gravity formalism, first proposed in 
\ct{einstein-23-1,einstein-23-2,eddington-book,schrodinger-book,kijowski-78}.
It has attracted since then a significant interest primarily due to the
established dynamical equivalence \ct{kijowski-82} of the three principal 
formulations of standard Einstein's gravity -- purely metric (second-order
formalism), metric-affine (Palatini or first-order formalism) and purely
affine formalism. For a more recent developments and list of references, see
\ct{Capozziello:2011et,poplawski-2014,Bejarano:2019zco,Delhom:2019zrb,Nascimento:2019qor,Delhom:2019btt,Olmo:2019qsj,Harko:2018gxr,Olmo:2017qab,Wojnar:2018hal,Afonso:2017aci,BeltranJimenez:2017doy,Casana:2015bea,Odintsov:2014yaa,Benisty:2018fgu,Benisty:2018ufz,Benisty:2018efx,Azri:2019ffj}, in particular about incorporating torsion and
explaining dark energy as an instrinsic property of space-time.

To establish the connection of our non-Riemannian volume-element
formalism and the purely affine formalism, our next task (Section 4) will be to 
represent the above quintessential inflationary model with a dynamical Higgs
effect in the form of a no-metric purely affine (Eddington-type) gravity.

\section{The Essence of the Non-Riemannian Volume-Form Formalism}

Volume-forms define volume-elements (generally covariant integration measures) 
over differentiable manifolds $\cM$, not necessarily Riemannian ones, so {\em no} metric 
is \textsl{a priori} needed \ct{spivak}. They are given by nonsingular maximal-rank 
differential forms $\om$ on $\cM$ (for definiteness we will consider the case of $D=4$ 
dimensional $\cM$):
\be
\int_{\cM} \om \bigl(\ldots\bigr) = \int_{\cM} d^4 x\, \Omega \bigl(\ldots\bigr) \;
\lab{omega-1}
\ee
where:
\be
\om = \frac{1}{4!}\om_{\m\n\k\l} dx^{\m}\wedge dx^{\n}\wedge dx^{\k}\wedge dx^{\l}
\quad ,\quad 
\om_{\m\n\k\l} = - \vareps_{\m\n\k\l} \Omega \;\; ,\;\;
\Omega = \frac{1}{4!}\vareps^{\m\n\k\l} \om_{\m\n\k\l} \; .
\lab{omega-2}
\ee
The conventions for the alternating symbols $\vareps^{\m\n\k\l}$ and
$\vareps_{\m\n\k\l}$ are: $\vareps^{0123}=1$ and $\vareps_{0123}=-1$.
The volume-element density (integration measure density)
$\Omega$ transforms as scalar density under general coordinate reparametrizations.

In standard general-relativistic theories the Riemannian spacetime volume-form 
is defined through the tetrad canonical one-forms 
$e^A = e^A_\m dx^\m$ ($A=0,1,2,3$):
\be
\om = e^0 \wedge e^1 \wedge e^2 \wedge e^3 = \det\Vert e^A_\m \Vert\,
dx^{0}\wedge dx^1 \wedge dx^2 \wedge dx^{3} \; ,
\lab{omega-riemannian-1}
\ee
which yields:
\be
\Omega = \det\Vert e^A_\m \Vert = \sqrt{-\det\Vert g_{\m\n}\Vert}
\equiv \sqrt{-g} \; .
\lab{omega-riemannian-2}
\ee
Instead of $\sqrt{-g}\, d^4 x$ we can employ another alternative {\em non-Riemannian} 
volume-element as in \rf{omega-1}-\rf{omega-2} given by a non-singular 
{\em exact} $4$-form $\om = d A$ where:
\br
A = \frac{1}{3!} A_{\m\n\k}
dx^{\m}\wedge dx^{\n}\wedge dx^{\k}  \quad \longrightarrow \quad
\om = \frac{1}{4!} \pa_{\lb \m} A_{\n\k\l \rb} 
dx^{\m}\wedge dx^{\n}\wedge dx^{\k}\wedge dx^{\l} \; .
\lab{Phi-4}
\er
Therefore, the corresponding non-Riemannian volume-element density
\be
\Omega \equiv \P(A) =
\frac{1}{3!}\vareps^{\m\n\k\l}\, \pa_{\m} A_{\n\k\l}.
\lab{omega-nonriemannian}
\ee
is defined in terms of
the dual field-strength scalar density of an auxiliary rank 3 tensor gauge field 
$A_{\m\n\k}$.

In the next Section we will discuss in some detail the properties of a
quintessential inflationary model coupled to a truncated version of the
electro-weak particle content carrying the standard electro-weak $SU(2)\times U(1)$
symmetry. Namely, for simplicity we retain only a Higgs-like scalar field
and discard the electro-weak gauge fields and fermions.

Before proceeding let us note the following important property of Lagrangian action
terms involving (one or more) non-Riemannian volume-elements:
\be
S = \int d^4 x \,\sum_j \P(A^{(j)})\, \cL^{(j)}({\rm other ~fields}) + \ldots \; .
\lab{NRVF-0}
\ee
The equations of motion of \rf{NRVF-0} w.r.t. the auxiliary tensor gauge fields 
$A^{(j)}_{\m\n\k}$ according to \rf{omega-nonriemannian} imply:
\be
\pa_\m \cL^{(j)}({\rm other ~fields}) = 0 \; \longrightarrow \; 
\cL^{(j)}({\rm other ~fields}) = M_j  \; ,
\lab{L-M}
\ee
where $M_j$ are {\em free integration constants} not present in the original
action \rf{NRVF-0}. This illustrates the significant advantage of the
non-Riemannian volume-element formalism over the ``Lagrange-multiplier gravity''
method \ct{lagrange-miltiplier-grav}, which appeared a decade later and which
requires picking \textsl{a priori} some {\em ad hoc} constant as opposed to
the dynamical appearance of the arbitrary integration constants \rf{L-M}.
For further advantages of the non-Riemannian volume-element formalism, see
the above remarks.

A characteristic feature of the modified gravitational theories \rf{NRVF-0} is that
when starting in the first-order (Palatini) formalism all non-Riemannian
volume-elements $\P(A^{(j)})$ yield almost {\em pure-gauge} degrees of freedom, 
\textsl{i.e.} they {\em do not} introduce any 
additional physical (field-propagating) gravitational degrees of freedom 
except for few discrete degrees 
of freedom with conserved canonical momenta appearing as arbitrary integration 
constants $M_j$. The reason is that the modified gravity action \rf{NRVF-0} in Palatini
formalism is linear w.r.t. the velocities of some of the components of the 
auxiliary gauge fields $A^{(j)}_{\m\n\k}$  defining the non-Riemannian 
volume-element densities, and does not depend on the velocities of the rest of 
auxiliary gauge field components. 
The (almost) pure-gauge nature of the latter is explicitly shown 
in   \ct{grav-bags,grf-essay} (appendices A) employing the standard 
canonical Hamiltonian treatment of systems with gauge symmetries, i.e.,
systems with first-class Hamiltonian constraints a'la Dirac 
\ct{henneaux-teitelboim,rothe}.

\section{Quintessential Inflationary Model with Dynamical Higgs Effect}

Our starting point is the following specific example of the general class of 
modified gravity models \ct{susyssb-1,grav-bags,grf-essay,varna-17,bpu-10,EPJC-79,NPB-951}) involving several non-Riemannian volume-elements (using units with $16\pi G_{\rm Newton}=1$):
\be
S = \int d^4 x\,\P_1 (A)\Bigl\lb R(g,\G) 
- 2 \L_0 \frac{\P_1 (A)}{\sqrt{-g}} + X_\phi + f_1 e^{\a\phi} + X_\s
- V_0 (\s) e^{\a\phi}\Bigr\rb + 
\int d^4 x\,\P_2 (B)\Bigl\lb f_2 e^{2\a\phi} - \frac{\P_0 (C)}{\sqrt{-g}}\Bigr\rb
\; .
\lab{NRVF-1}
\ee
Here the following notations are used:
\begin{itemize}
\item
The scalar curvature $R(g,\G) = g^{\m\n} R_{\m\n}(\G)$ is given in terms
of the Ricci tensor $R_{\m\n}(\G)$ in the first-order (Palatini) formalism:
\be
R_{\m\n}(\G) = \pa_\a \G^\a_{\m\n} - \pa_\n \G^\a_{\m\a}
+\G^\a_{\a\b} \G^\b_{\m\n} - \G^\a_{\b\n} \G^\b_{\m\a}
\lab{Ricci}
\ee
defined by the affine connection $\G^\l_{\m\n}$ \textsl{a priori} 
independent of the metric $g_{\m\n}$.
\item
The non-Riemannian volume-element densities $\P_1 (A), \P_2 (B), \P_0 (C)$
are defined as in \rf{omega-nonriemannian}:
\be
\P_1 (A) = \frac{1}{3!}\vareps^{\m\n\k\l}\, \pa_{\m} A_{\n\k\l} \quad,\quad
\P_2 (B) = \frac{1}{3!}\vareps^{\m\n\k\l}\, \pa_{\m} B_{\n\k\l} \quad,\quad
\P_0 (C) = \frac{1}{3!}\vareps^{\m\n\k\l}\, \pa_{\m} C_{\n\k\l} \; .
\lab{omega-nonriemannian-1}
\ee
\item
$\phi$ is a neutral scalar ``inflaton'' and $\s \equiv (\s_a)$ is a complex 
$SU(2)\times U(1)$ iso-doublet Higgs-like scalar field
with the isospinor index $a=+,0$ indicating the corresponding $U(1)$ charge.
The corresponding kinetic energy terms in \rf{NRVF-1} read:
\be
X_\phi \equiv - \h g^{\m\n} \pa_\m \phi \pa_\n \phi \quad ,\quad
X_\s \equiv - g^{\m\n} \pa_\m \s^{*}_a \pa_\n \s_a \; ,
\lab{short}
\ee
and
\be
V_0 (\s) \equiv m_0^2\, \s^{*}_a \s_a \; ,
\lab{mass-term}
\ee
is a canonical mass term for the Higgs-like field, \textsl{i.e.}, neither
negative (``ghost-like'') mass-squared term nor quartic self-interaction are
introduced unlike the case in the standard electro-weak model 
\ct{standard-particle-model-1,standard-particle-model-2}.
\item
$f_{1,2}$ and $\a$ are dimensionful coupling constants in the ``inflaton''
potential. $\L_0$ is small dimensionful constant which will be identified in
the sequel with the ``late'' universe cosmological constant in the dark
energy dominated accelerated expansion's epoch.
\end{itemize}

The specific form of the action \rf{NRVF-1} is fixed by the requirement of global
Weyl-scale invariance  under:
\br
g_{\m\n} \to \l g_{\m\n} \;, \; 
A_{\m\n\k} \to \l A_{\m\n\k} \; ,\; B_{\m\n\k} \to \l^2 B_{\m\n\k} \; ,\; 
C_{\m\n\k} \to C_{\m\n\k} \; ,
\lab{scale-transf-1} \\
\phi \to \phi - \frac{1}{\a}\ln\l \quad ,\quad \s_a \to \s_a  \; ,
\lab{scale-transf-2}
\er
where scaling parameter $\lambda = {\rm const}$. 
The importance of global scale symmetry within the 
context of non-Riemannian volume-element formalism has been already  stressed in 
the first original papers (see \ct{TMT-orig-1}), where in particular 
models with spontaneously broken dilatation symmetry have been constructed 
along these lines, which are free of the Fifth Force Problem \ct{Guendelman:2007ph}.

Varying the action \rf{NRVF-1} w.r.t. $g^{\m\n}$, $\G^\l_{\m\n}$,
$A_{\m\n\l}$, $B_{\m\n\l}$, $C_{\m\n\l}$, $\phi$ and $\s^a$
, yield the following equations of motion, respectively:
\br
R_{\m\n}(\G) - \L_0 \frac{\P_1(A)}{\sqrt{-g}} g_{\m\n} 
- \h \pa_\m \phi \pa_\n \phi -  \pa_\m \s^{*}_a \pa_\n \s_a
-\h g_{\m\n} \frac{\P_2(B)\P_0(C)}{\P_1(A)\sqrt{-g}} = 0 \; ,
\lab{einstein-like}\\
\P_1(A) g^{\m\n}
\Bigl(\nabla_\l \d\G^{\l}_{\m\n} - \nabla_\m \d\G^{\l}_{\l\n}\Bigr) = 0 \; ,
\lab{var-G} \\
g^{\m\n}\Bigl(R_{\m\n}(\G) - \h \pa_\m \phi \pa_\n \phi
- \pa_\m \s^{*}_a \pa_\n \s_a \Bigr) - 4\L_0 \frac{\P_1(A)}{\sqrt{-g}}
+ \bigl(f_1 - m^2_0\, \s^{*}_a \s_a\bigr) e^{\a\phi} = M_1 \equiv {\rm const} \; ,
\lab{M1} \\
f_2 e^{-2\a\phi} - \frac{\P_0 (C)}{\sqrt{-g}} = - M_2 \equiv {\rm const} \quad ,\quad
\frac{\P_2 (B)}{\sqrt{-g}} = \chi_2 \equiv {\rm const} \; ,
\lab{M2-chi2} \\
\pa_\m \Bigl(\P_1(A)g^{\m\n}\pa_\n\phi\Bigr) 
+ \a \P_1(A)\bigl( f_1 - m^2_0\, \s^{*}_a \s_a\bigr) e^{\a\phi}
+ 2\a \P_2(B) f_2 e^{2\a\phi} = 0 \; ,
\lab{phi-eq}\\
\pa_\m \Bigl(\P_1(A)g^{\m\n}\pa_\n\s_a\Bigr) - \P_1(A) m^2_0\, e^{\a\phi} \s_a =0 \; . 
\lab{higgs-eq}
\er
Eqs.\rf{M1}-\rf{M2-chi2} are special cases of the general Eq.\rf{L-M} discussed
above. Here $M_{1,2}$ and $\chi_2$ are arbitrary (dimensionful and dimensionless, 
respectively) integration constants, with $M_{1,2}$ triggering a spontaneous 
breaking of the global Weyl-scale symmetry \rf{scale-transf-2}.

Taking the trace of Eq.\rf{einstein-like} and comparing with \rf{M1}-\rf{M2-chi2}
we find for the ratio of volume-element densities:
\be
\chi_1 \equiv \frac{\P_1(A)}{\sqrt{-g}}= 
\frac{2\chi_2\bigl(f_2 e^{2\a\phi}+M_2\bigr)}{M_1 +
\bigl(m^2_0\, \s^{*}_a \s_a - f_1\bigr) e^{\a\phi}} \equiv \chi_1 (\phi,\s) \; .
\lab{chi-1}
\ee

On the other hand, following analogous derivation in  \ct{TMT-orig-1}, 
Eq.\rf{var-G} yields a solution for $\G^\m_{\n\l}$ as a Levi-Civita
connection:
\be
\G^\m_{\n\l} = \G^\m_{\n\l}({\bar g}) =
\h {\bar g}^{\m\k}\(\pa_\n {\bar g}_{\l\k} + \pa_\l {\bar g}_{\n\k}
- \pa_\k {\bar g}_{\n\l}\)
\lab{Levi-Civita-bar}
\ee
w.r.t. to a {\em Weyl-conformally rescaled} metric: 
\be
{\bar g}_{\m\n} = \chi_1 (\phi,\s)\, g_{\m\n} 
\lab{g-bar}
\ee
with $\chi_1 (\phi,\s)$ as in \rf{chi-1}.

Conformal transformation $g_{\m\n} \to {\bar g}_{\m\n}$ via \rf{g-bar}
convert the modified gravity action \rf{NRVF-1} into the physical
Einstein-frame action (objects in the Einstein-frame indicated by a bar):
\be
S_{\rm EF} = \int d^4 x\,\sqrt{-{\bar g}} \Bigl\lb R({\bar g})
- \h {\bar g}^{\m\n} \pa_\m \phi \pa_\n \phi 
- {\bar g}^{\m\n} \pa_\m \s^{*}_a \pa_\n \s_a - U_{\rm eff}(\phi,\s)\Bigr\rb \; ,
\lab{NRVF-EF}
\ee
with an effective Einstein-frame scalar field potential:
\br
U_{\rm eff}(\phi,\s) \equiv 
\frac{M_1 + e^{\a\phi}\bigl(m^2_0\, \s^{*}_a \s_a - f_1\bigr)}{\chi_1 (\phi,\s)}
- \frac{\chi_2 \bigl(f_2 e^{2\a\phi} + M_2\bigr)}{\bigl(\chi_1 (\phi,\s)\bigr)^2}
+ 2 \L_0
\nonu \\
= \frac{\Bigl\lb M_1 + e^{\a\phi}\bigl(m^2_0\, \s^{*}_a \s_a 
- f_1\bigr)\Bigr\rb^2}{4\chi_2 \bigl(f_2 e^{2\a\phi}+M_2\bigr)} + 2\L_0 \; ,
\lab{U-eff}
\er
which is entirely {\em dynamically generated} due to the appearance of the
free integration constants $M_{1,2}$ and $\chi_2$ \rf{M1}-\rf{M2-chi2}.

As discussed in \ct{Guendelman:2014bva,grav-bags,varna-17,bpu-10} 
the scalar potential $U_{\rm eff}(\phi,\s)$ \rf{U-eff} has a remarkable
feature -- it possesses two (infinitely) large flat regions
as a function of $\phi$ at $\s_a = {\rm fixed}$ (see the graphical representation on 
Fig.1) with the following properties:

\begin{figure}[h]
\begin{center}
\includegraphics[width=12cm,keepaspectratio=true]{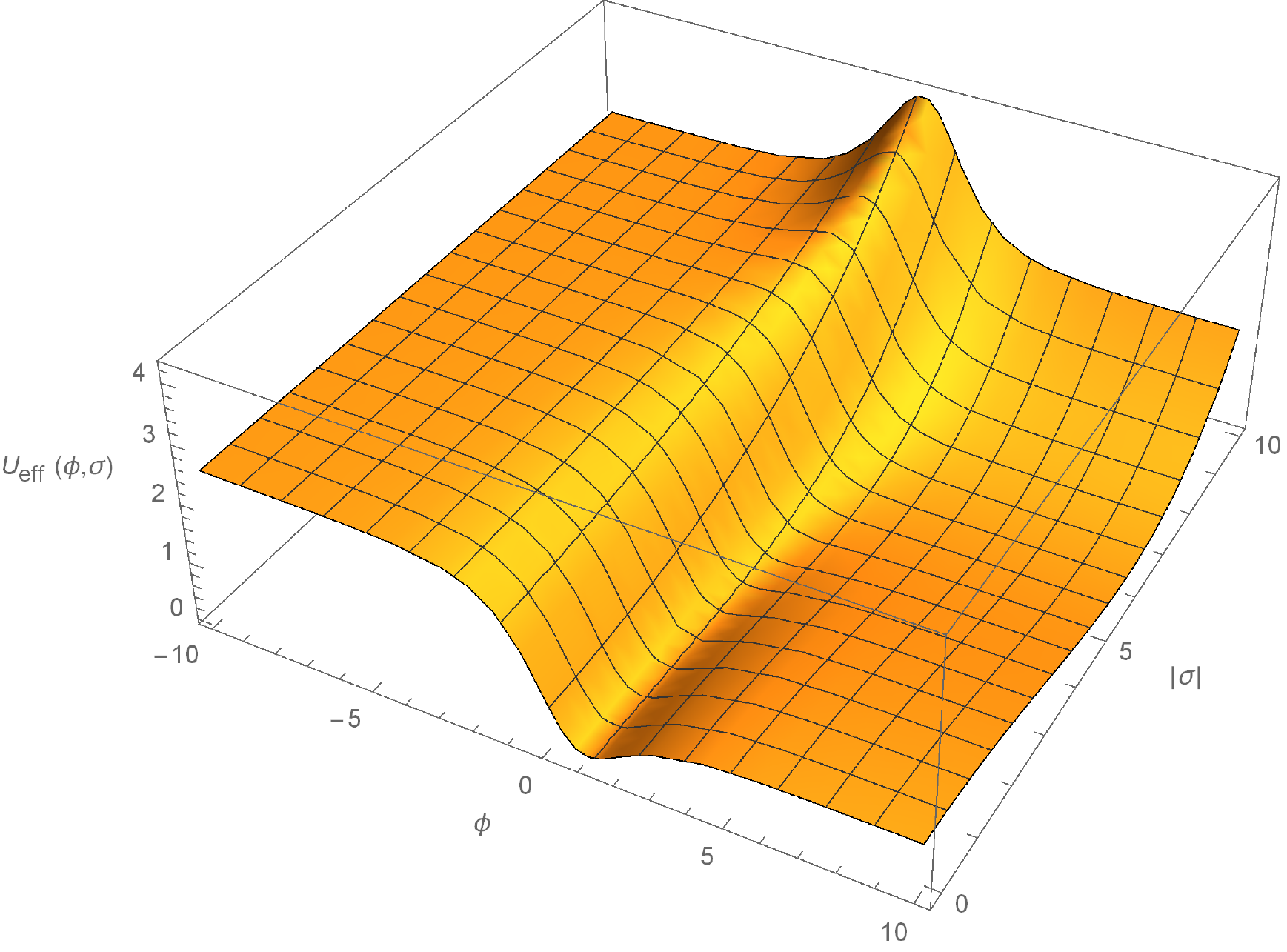}
\caption{Qualitative shape of the two-dimensional plot for 
the effective scalar potential $U_{\rm eff}(\phi,\s)$ \rf{U-eff}.}
\end{center}
\end{figure}

\begin{itemize}
\item
(a) (-) flat ``inflaton'' region for large negative values of $\phi$ (and
$\s_a$ -- finite)
corresponding to the ``slow-roll'' inflationary evolution of the ``early''
universe driven by $\phi$. 
Here the effective potential \rf{U-eff} reduces to (an almost) constant value
independent of the finite value of $\s_a$ -- this is energy scale of the 
inflationary epoch: 
\be
U_{\rm eff}\bigl(\phi,\s\bigr) \simeq U_{(-)} = 
\frac{M_1^2}{4\chi_2\,M_2} + 2\L_0 \; .
\lab{U-minus}
\ee
Thus, in the ``early'' universe the Higgs-like field $\s_a$ must be (approximately)
either massless or constant with {\em no} non-zero vacuum expectation value, 
therefore there is {\em no spontaneous breaking} of $SU(2)\times U(1)$ symmetry.
Moreover, in fact as shown in the Remark below, $\s_a$ does not participate in 
the ``slow-roll'' inflationary evolution, so $\s$ stays constant there equal to the
``false''vacuum value $\s=0$.
\item
(b) (+) flat ``inflaton'' region for large positive values of $\phi$ (and
$\s_a$ -- finite) corresponding to the evolution of the post-inflationary
(``late'') universe, where:
\be
U_{\rm eff}\bigl(\phi,\s\bigr) \simeq U_{(+)}(\s)
=\frac{\Bigl( m^2_0\, \s^{*}_a \s_a - f_1\Bigr)^2}{4\chi_2 f_2} + 2\L_0 
\lab{higgs-potential}
\ee
acquires the form of a dynamically induced $SU(2)\times U(1)$ spontaneous
symmetry breaking Higgs potential with a Higgs ``vacuum'' at:
\be
|\s_{\rm vac}|=\frac{1}{m_0}\sqrt{f_1} \; ,
\lab{sigma-vac}
\ee
where the parameters are naturally identified as:
\be
f_1 \sim M^4_{EW} \quad ,\quad m_0 \sim M_{EW}
\lab{f1-m0-scales}
\ee
in terms of the electro-weak energy scale $M_{EW} \sim 10^{-16} M_{Pl}$.
\item
Thus, the residual cosmological
constant $\Lambda_0$ in \rf{higgs-potential} has to be identified with the current 
epoch observable cosmological constant ($\sim 10^{-122} M^4_{Pl}$) and, therefore, 
according to \rf{U-minus} the integration constants $M_{1,2}$ are naturally 
identified by orders of magnitude as
\be
M_1 \sim M_2 \sim 10^{-8} M_{Pl}^4 \; ,
\lab{M12-scales}
\ee
since the latter case the order of magnitude 
of the vacuum energy density in the {\em (-) flat region} \rf{U-minus} becomes:
\be
U_{(-)} \sim M_1^2/M_2 \sim 10^{-8} M_{Pl}^4 \; ,
\lab{U-minus-magnitude}
\ee
which conforms to the Planck Collaboration data \ct{Planck-1,Planck-2} for the
``early'' universe's energy scale of inflation being of order $10^{-2} M_{Pl}$.
\item
Here the order of magnitude for $f_2$ is determined from the mass term of the
Higgs-like field $\s$ in the $(+)$ flat region resulting from \rf{higgs-potential} 
upon expansion around the Higgs vacuum 
($\s = \s_{\rm vac} + {\wti \s}$):
\be
\frac{f_1 m_0^2}{\chi_2 f_2}\, ({\wti \s}_a)^{*} ({\wti \s}_a) \; ,
\lab{Higgs-mass-term}
\ee
which implies that:
\be
f_2 \sim f_1 \sim M^4_{EW} \; .
\lab{f2-scale}
\ee
\item
Let us specifically note that the viability of the present model 
(in a slightly simplified form without the Higgs scalar) concerning confrontation 
with the observational data has already been analyzed and confirmed
numerically in Ref.\ct{Guendelman:2014bva}.
In particular, a graphical plot of the evolution of $r$
(tensor-to-scalar ratio) vs $n_s$ (scalar spectral index) has been provided there.
\end{itemize}

\textbf{Remark.}
Assuming that in the (-) flat ``inflaton'' region (for large negative values of 
$\phi$ and $\s_a$ -- finite) both the ``inflaton'' $\phi$ and the Higgs-like
field $\s_a$ evolve in a
``slow-role'' regime, their ``slow-role'' equations of motion in the
standard FLRW reduction of the Einstein-frame metric 
(${\bar g}_{\m\n}dx^\m dx^\n \equiv - N^2(t) dt^2 + a^2(t) d{\vec x}.d{\vec x}$)
read accordingly (cf. \textsl{e.g.} 
  \ct{slow-roll-1,slow-roll-2,slow-roll-param-1,slow-roll-param-2}):
\br
\phidot\, \simeq - \frac{1}{3H} \partder{U_{\rm eff}(\phi,\s)}{\phi} ,\quad
\partder{U_{\rm eff}(\phi,\s)}{\phi} = \frac{\a e^{\a\phi}
\Bigl\lb M_1 + e^{\a\phi}\bigl(m^2_0\,|\s|^2 - f_1\bigr)\Bigr\rb\, 
\Bigl\lb M_2 \bigl(m^2_0\,|\s|^2 - f_1\bigr) - 
M_1 f_2 e^{\a\phi}\Bigr\rb}{2\chi_2 \bigl(M_2 + f_2 e^{2\a\phi}\bigr)^2}
\lab{slow-roll-phi} \\
\sdot\, \simeq - \frac{1}{3H} \partder{U_{\rm eff}(\phi,\s)}{\s} \quad 
\longrightarrow \quad
\frac{d|\s|}{dt} \simeq - \frac{1}{3H}
\frac{m^2_0\,|\s| e^{\a\phi} 
\Bigl\lb M_1 + e^{\a\phi}\bigl(m^2_0\,|\s|^2 - f_1\bigr)\Bigr\rb}{
2\chi_2 \bigl(M_2 + f_2 e^{2\a\phi}\bigr)} \; ,
\lab{slow-roll-sigma}
\er
where $|\s|^2 \equiv \s^{*}_a \s_a$ and $H=\frac{\adot}{a}$ denotes the Hubble 
parameter. Eqs.\rf{slow-roll-phi}-\rf{slow-roll-sigma} define parametrically 
a curve $|\s| = |\s| (\phi)$ in the two-field $(\phi,|\s|)$ target space.
Equivalently, this curve is defined through the differential equation:
\be
\frac{d|\s|}{dz}\,\simeq \frac{m^2_0\,|\s(z)| \Bigl(M_2 + f_2 z^2\Bigr)}{\a^2\,z\,
\Bigl\lb M_2 \bigl(m^2_0 |\s(z)|^2 - f_1\bigr) - M_1 f_2 z\Bigr\rb}
\quad ,\quad z\equiv e^{\a\phi} \; .
\lab{sigma-curve}
\ee
In the (-) flat ``inflaton'' region ($\phi$ -- large negative) $z$ is very
small, so in this case Eq.\rf{sigma-curve} can be rewritten as:
\be
\a\Bigl(|\s| - \frac{f_1}{m^2_0\,|\s|}\Bigr) d|\s| \simeq \frac{dz}{z} \quad
\longrightarrow \quad 
\a\Bigl(\h|\s(z)|^2 - \frac{f_1}{m^2_0} \ln|\s(z)|\Bigr) \simeq \ln z \; .
\lab{sigma-inconsist}
\ee
Obviously, a consistent solution $|\s(z)|$ of \rf{sigma-inconsist} {\em does not} 
exist for $z=e^{\a\phi} \to 0$, therefore, the assumption for the ``slow-roll''
evolution \rf{slow-roll-sigma} of the Higgs-like field $\s_a$ in the inflationary region 
(large negative values of $\phi$) is invalid. Thus $|\s|$ must be constant and 
Eq.\rf{slow-roll-sigma} implies $|\s|=0$ in the (-) flat ``inflaton'' region.

\bigskip

To conclude this section, we see that thanks to the remarkable dynamically
generated scalar potential \rf{U-eff} the ``inflaton'' $\phi$ plays the role both of
driving ``slow-roll'' inflationary dynamics in the ``early'' universe, as well
as it plays the role of a quintessential variable dark-energy field triggering slowly
accelerating de Sitter expansion in the ``late'' universe.

Accordingly, gravity-inflaton dynamics generates dynamically spontaneous 
$SU(2)\times U(1)$ symmetry breaking -- Higgs effect -- in the
post-inflationary epoch, whereas it dynamically suppresses spontaneous symmetry breaking
during inflation 
in the ``early'' universe.
Namely, the dynamical transition from the ``false'' Higgs vacuum to the genuine
electroweak spontaneously broken vacuum is driven by the inflaton
evolving from (large) negative values (on the ``(-)'' flat region
\rf{U-minus} of the scalar potential \rf{U-eff}) to (large) positive values 
(on the ``(+)'' flat region \rf{higgs-potential} of the scalar potential \rf{U-eff}).

Thus, our scale invariant modified gravity model \rf{NRVF-1} in 
its Einstein-frame representation \rf{NRVF-EF}-\rf{U-eff} turns out to be an explicit 
implementation of Bekenstein's idea \ct{bekenstein-86} about a gravity-assisted 
spontaneous symmetry breaking of electro-weak (Higgs) type without invoking 
negative mass squared and a quartic Higgs field
self-interaction unlike the canonical case in the standard particle model
\ct{standard-particle-model-1,standard-particle-model-2}.

\section{Eddinton-type No-Metric Gravity and Quintessential Inflation}

Let us now consider a generic model of gravity, with some Riemannian
metric ${\bar g}_{\m\n}$ and with the ordinary Riemannian
volume-element $\sqrt{-{\bar g}}$ within the first-order (Palatini)
formalism, interacting with a multi-component scalar field 
$\phi^A$, $A=1,\ldots, N$ (using again units with 
$16\pi G_{\rm Newton}=1$):
\be
S = \int d^4 x\,\sqrt{-{\bar g}}\Bigl\lb {\bar g}^{\m\n} R_{\m\n}(\G) 
- \h {\bar g}^{\m\n} h_{AB} \pa_\m \phi^A \pa_\n \phi^B - U (\phi)\Bigr\rb \; ,
\lab{action-0}
\ee
where the Ricci tensor $R_{\m\n}(\G)$ is the same as in \rf{Ricci}, and
$h_{AB}(\phi)$ indicates some ``metric'' in the scalar field target space
(in the present case it will be just a unit matrix).

The equations of motion w.r.t $g^{\m\n}$, $\phi^A$ and $\G^\l_{\m\n}$ read
accordingly:
\br
R_{\m\n}(\G) = \h \Bigl( T_{\m\n} - \h {\bar g}_{\m\n} T^\l_\l\Bigr) \; ,
\lab{einstein-1}\\
T_{\m\n} = h_{AB}(\phi) \pa_\m \phi^A \pa_\n \phi^B 
- {\bar g}_{\m\n} \Bigl\lb \h {\bar g}^{\k\l} h_{AB}(\phi) \pa_\k \phi^A \pa_\l \phi^B 
+ U (\phi)\Bigr\rb \; ,
\lab{EM-1}
\er
\br
\frac{1}{\sqrt{-{\bar g}}}\pa_\m \bigl(
\sqrt{-{\bar g}} g^{\m\n} h_{AB} \pa_\n \phi^B\bigr)
- \h {\bar g}^{\m\n} \partder{h_{CD}}{\phi^A} \pa_\m \phi^C \pa_\n \phi^D
- \partder{U(\phi)}{\phi^a} = 0 \; ,
\lab{phi-eq-0} \\
\int d^4 x\,\sqrt{-{\bar g}} {\bar g}^{\m\n} 
\Bigl(\nabla_\l \d\G^{\l}_{\m\n} - \nabla_\m \d\G^{\l}_{\l\n}\Bigr) = 0
\lab{var-G-0}
\er
Following again the analogous derivation in \ct{TMT-orig-1}, the solution of 
Eq.\rf{var-G-0} is 
that $\G^{\l}_{\m\n}$ becomes the canonical Levi-Civita connection w.r.t. 
${\bar g}_{\m\n}$:
\be
\G^\m_{\n\l} = \G^\m_{\n\l}({\bar g}) = 
\h {\bar g}^{\m\k}\(\pa_\n {\bar g}_{\l\k} + \pa_\l {\bar g}_{\n\k} 
- \pa_\k {\bar g}_{\n\l}\) \; .
\lab{G-eq}
\ee

Eqs.\rf{einstein-1}-\rf{EM-1} can be equivalently written as:
\be
{\bar g}_{\m\n} = \frac{2}{U(\phi)} 
\Bigl( R_{\m\n}(\G) - \h h_{AB}(\phi) \pa_\m \phi^A \pa_\n \phi^B\Bigr) \; ,
\lab{einstein-2}
\ee
that is, the metric ${\bar g}_{\m\n}$ in \rf{action-0} is expressed entirely in 
terms of the affine connection and the matter field.

Now, we will show that the gravity-matter theory \rf{action-0} is
equivalent, in a sense of producing the same equations of motion
\rf{EM-1}-\rf{G-eq}, to the following Eddington-type 
purely affine gravity theory:
\be
S_{Edd} = \int d^4 x \,\frac{2}{U(\phi)}
\sqrt{\det\Vert R_{\m\n}(\G) - \h h_{AB}(\phi) \pa_\m \phi^A \pa_\n \phi^B\Vert} \; ,
\lab{Edd-gravity}
\ee
\textsl{i.e.} \rf{Edd-gravity} does not involve at all a Riemannian metric.

Indeed, varying the action \rf{Edd-gravity} w.r.t. $\G^\l_{\m\n}$ and $\phi^a$ we get:
\be
\frac{2}{U(\phi)}\sqrt{\det\Vert H_{\a\b}(\G,\phi,\s)\Vert}
\bigl(H^{-1}((\G,\phi,\s)\bigr)^{\m\n} 
\Bigl(\nabla_\l \d\G^{\l}_{\m\n} - \nabla_\m \d\G^{\l}_{\l\n}\Bigr) = 0 
\; ,
\lab{var-G-Edd}
\ee
with the short-hand notation:
\be
H_{\m\n}(\G,\phi) \equiv 
R_{\m\n}(\G) - \h h_{AB}(\phi)\pa_\m \phi^A \pa_\n \phi^B \; ,
\lab{H-short}
\ee
and
\br
\pa_\m \Bigl(\frac{1}{U(\phi)}\sqrt{\det\Vert H_{\a\b}(\G,\phi)\Vert}
\bigl(H^{-1}(\G,\phi)\bigr)^{\m\n} h_{AB}(\phi) \pa_\n \phi^B \Bigr) 
\nonu \\
-\frac{1}{2U(\phi)}\sqrt{\det\Vert H_{\a\b}(\G,\phi)\Vert}
\bigl(H^{-1}(\G,\phi)\bigr)^{\m\n} \partder{h_{CD}}{\phi^A}
\pa_\m \phi^C \pa_\n \phi^D
+ 2\partder{}{\phi}\Bigl(\frac{1}{U(\phi)}\Bigr)\,
\sqrt{\det\Vert H_{\a\b}(\G,\phi)\Vert} = 0 \; .
\lab{var-phi-Edd}
\er

Now, using the identification Eq.\rf{einstein-2} for the Riemannian metric
${\bar g}_{\m\n} = \frac{2}{U(\phi)} H_{\m\n}(\G,\phi)$ with $H_{\m\n}(\G,\phi)$ as
in \rf{H-short}, Eqs.\rf{var-G-Edd}-\rf{var-phi-Edd} become identical to 
Eqs.\rf{var-G-0} and \rf{phi-eq-0}, respectively. 

The above derivation of
purely affine gravity interacting with multi-component scalar fields appeared
previously in  \ct{azri-ijmpd}. Historically, this formulation was proposed for the
first time in  \ct{kijowski-78} in the special case of a single Klein-Gordon field 
with $U(\phi) = \h m^2\,\phi^2$, see also \ct{kijowski-07}.

Applying the above established equivalence between the models \rf{action-0} and 
\rf{Edd-gravity} to the initial modified gravity action \rf{NRVF-1} and its
Einstein-frame representation \rf{NRVF-EF} with $U_{\rm eff}(\phi,\s)$ as in 
\rf{U-eff}, analyzed in Section 3 above, we find that the following specific 
Eddington-type purely affine no-metric gravity model:
\br
S_{Edd} = \int d^4 x \,
\frac{8\chi_2\Bigl(f_2 e^{-2\a\phi} + M_2\Bigr)}{\Bigl\lb M_1 +
\bigl(m_0^2\,\s^{*}_a \s_a - f_1 e^{-\a\phi}\bigr)\Bigr\rb^2
+ 8\L_0 \chi_2 \bigl(f_2 e^{-2\a\phi} + M_2\bigr)} 
\nonu \\
\times \sqrt{\det\Vert R_{\m\n}(\G) - \h \pa_\m \phi \pa_\n \phi
- \pa_\m \s^{*}_a \pa_\n \s_a\Vert} \; ,
\lab{Edd-gravity-1}
\er
actually describes a quintessential inflationary dynamics with dynamically
generated Higgs effect in the post-iflationary epoch with all the
properties discussed in section 3. The metric $g_{\m\n}$ in the initial modified
scale-invariant gravity action \rf{NRVF-1} with non-Riemannian
volume-elements, taking into account relation \rf{einstein-2}
applied for the Einstein-frame metric \rf{g-bar} where 
$U(\phi) = U_{\rm eff}(\phi,\s)$ \rf{U-eff} and using the on-shell relations
\rf{M2-chi2} and \rf{chi-1}, is identified as:
\be
g_{\m\n} = \frac{2}{\chi_1(\phi,\s)\,U_{\rm eff}(\phi,\s)} 
\Bigl\lb R_{\m\n}(\G) - \h \pa_\m \phi \pa_\n \phi
- \pa_\m \s^{*}_a \pa_\n \s_a\Bigr\rb\; ,
\lab{metric-NRVF}
\ee
with $R_{\m\n}(\G)$ as in \rf{Ricci}, and $\chi_1(\phi,\s)$ and $U_{\rm eff}(\phi,\s)$
explicitly given in \rf{chi-1} and \rf{U-eff}, respectively:
\be
\frac{1}{\chi_1(\phi,\s)\,U_{\rm eff}(\phi,\s)} =
\frac{2\Bigl\lb M_1 + e^{\a\phi}\bigl(m^2_0\,\s^{*}_a \s_a - f_1\bigr)\Bigr\rb}{
\Bigl\lb M_1 + e^{\a\phi}\bigl(m^2_0\,\s^{*}_a \s_a - f_1\bigr)\Bigr\rb^2 
+ 8\L_0 \chi_2 \Bigl(f_2 e^{2\a\phi} + M_2\Bigr)} \; .
\lab{chi1-U}
\ee
\section{Conclusions}

In the present paper we have employed two fundamental concepts -- of non-Riemannian 
metric-independent spacetime volume-elements and of (global) scale
invariance to construct a self-consistent model of modified gravity coupled
to a neutral scalar ``inflaton'' and to a Higgs-like $SU(2)\times U(1)$
iso-boublet scalar possessing the following extraordinary features: 

$\phantom{aaa}$(a) In the
physical Einstein frame, thanks to a dynamical generation of a remarkable
scalar potential with two long flat ``inflaton'' regions with vastly different
heights, the model describes a plausible quintessential inflationary scenario,
driven by the ``inflaton'', with a ``slow-roll'' inflationary stage in the ``early'' 
universe and a slow accelerating de Sitter expansion in the ``late'' universe;

$\phantom{aaa}$(b) This model provides an explanation of the interplay
between cosmological dynamics and the patterns of symmetry breaking during
the evolution of the universe. Namely, we find an explicit realization from first
(Lagrangian-action) principles of the noteworthy proposal of Bekenstein from 1986 
about ``gravity-assisted'' dynamical Higgs-like spontaneous symmetry breakdown
(Higgs effect). We exhibit gravity-``inflaton'' suppression of the
Higgs effect during inflation,\textsl{i.e.}, no electroweak spontaneous
breakdown there), whereas in the post-inflationary epoch a
Higgs-type symmetry breaking potential is dynamically created. 

$\phantom{aaa}$(c) The coupling constants in the initial modified gravity
action are naturally identified as powers of the standard electroweak mass scale.

$\phantom{aaa}$(d) It is shown how to represent the above quintessential
inflationary model with a dynamical Higgs effect in the form of a no-metric
purely affine (Eddington-type) gravity.


A next important task is to study in some detai, within the present quintessential
inflationary scenario with a dynamical Higgs effect, 
the numerical solutions for the basic inflationary observables 
(scalar power spectral index, tensor-to-scalar ratio, {\em etc.})
extending the numerical analysis from Ref.\ct{Guendelman:2014bva} 
(where the Higgs field was absent).
Since in the present scenario the Higgs field during (most of the) inflation
resides in its ``false'' vacuum, the significant impact of Higgs field
dynamics will occur after end of inflation when the ``inflaton'' starts to
generate the non-trivial Higgs symmetry breaking potential.

\bigskip
\acknowledgments

We all are grateful for support by COST Action CA-15117 (CANTATA) and 
COST Action CA-18108. D.B. thanks Ben-Gurion University of the Negev and 
Frankfurt Institute for Advanced Studies for generous support. 
E.N. and S.P. are partially supported by Bulgarian National Science Fund Grant DN 18/1.
Also we thank two of the referees for their constructive remarks which contributed
to the improvement of the presentation.

\bibliography{ref}

\end{document}